\documentclass[sigconf, nonacm]{acmart}

\newcommand\vldbdoi{XX.XX/XXX.XX}

\usepackage{hyperref}
\usepackage{listings}
\usepackage{graphicx}
\usepackage{tikz}
\usepackage{xcolor}
\usepackage{tcolorbox}
\usetikzlibrary{positioning, shapes, shadows, calc, backgrounds, fit, shapes.multipart, arrows.meta, shapes.geometric}

\newcommand{\xhdr}[1]{\vspace{1mm}\noindent{{\bf #1.}}}
\definecolor{lightblue}{RGB}{204, 229, 255}
\definecolor{lightred}{RGB}{255, 204, 204}
\definecolor{lightyellow}{RGB}{255, 247, 188}
\definecolor{lightgreen}{RGB}{204, 255, 204}
\definecolor{predict}{RGB}{78, 150, 198}
\definecolor{predictlight}{RGB}{242, 249, 252}
\definecolor{foreach}{RGB}{233, 137, 56}
\definecolor{foreachlight}{RGB}{252, 243, 236}
\definecolor{where}{RGB}{66, 126, 79}
\definecolor{aggr}{RGB}{165, 91, 215}
\tikzstyle{arrow} = [->, thick, >=stealth]

\begin{document}

\title{Predictive Query Language: A Domain-Specific Language for Predictive Modeling on Relational Databases}

\author{Vid Kocijan}
\affiliation{%
  \institution{NVIDIA\textsuperscript{$\dagger$}}
}

\author{Jinu Sunil}
\affiliation{%
  \institution{NVIDIA\textsuperscript{$\dagger$}}
  }

\author{Jan Eric Lenssen}
\affiliation{%
 \institution{MPI for Informatics\textsuperscript{$\dagger$}}
  }

\author{Viman Deb}
\affiliation{%
  \institution{Flowspec\textsuperscript{$\dagger$}}
}

\author{Xinwei He}
\affiliation{%
  \institution{Traceroot.ai\textsuperscript{$\dagger$}}
  }

\author{Federico Reyes Gomez}
\affiliation{%
  \institution{Tesora\textsuperscript{$\dagger$}}
  }

\author{Matthias Fey}
\affiliation{%
  \institution{NVIDIA\textsuperscript{$\dagger$}}
  }

\author{Jure Leskovec}
\affiliation{
  \institution{NVIDIA, Stanford University\textsuperscript{$\dagger$}}
  }

\begin{abstract}
The purpose of predictive modeling on relational data is to predict future or missing values in a relational database, for example, future purchases of a user, risk of readmission of the patient, or the likelihood that a financial transaction is fraudulent.
 Typically powered by machine learning methods, predictive models are used in recommendations, financial fraud detection, supply chain optimization, and other systems, providing billions of predictions every day.
 However, training a machine learning model requires manual work to extract the required training examples---prediction entities and target labels---from the database, which is slow, laborious, and prone to mistakes.
 
Here, we present the Predictive Query Language (PQL), an SQL-inspired declarative language for defining predictive tasks on relational databases.
PQL allows specifying a predictive task in a single declarative query, enabling the automatic computation of training labels for a large variety of machine learning tasks, such as regression, classification, time-series forecasting, and recommender systems.
PQL is already successfully integrated and used in a collection of use cases as part of a predictive AI platform. The versatility of the language can be demonstrated through its many ongoing use cases, including financial fraud, item recommendations, and workload prediction. We demonstrate its versatile design through two implementations; one for small-scale, low-latency use, and one that can handle large-scale databases. 

%
\end{abstract}

\maketitle

\begingroup
\renewcommand\thefootnote{}\footnote{\noindent
$\dagger$ Work conducted under Kumo.ai affiliation, now part of NVIDIA.
}\addtocounter{footnote}{-1}\endgroup

\section{Introduction}
Creating a new machine learning model based on data from a relational database traditionally requires a large engineering effort.
This is due to the inherently iterative nature of ML model development, where slow iteration dramatically lengthens development cycles, driving up costs through prolonged use of scarce expert labor and expensive computing resources. It is therefore highly desirable to accelerate the model development process through simplification and automation.


A typical machine learning development cycle on relational databases comes with three major bottlenecks: feature engineering, training label/table generation, and model training.
Input feature engineering and model training have received much attention from the research community, with recent advances in relational deep learning~\cite{rdl, ranjan2025relational, dwivedi2025relgt}, foundation models~\cite{hollmann2025tabpfn, qu2025tabicl, kumorfm}, and in-context learning~\cite{muller2022transformers}, which significantly reduce or even eliminate the need for them. However, generating training tables with training entities and their corresponding tables received much less attention.

\begin{figure}[t]
    \centering
    \includegraphics[width=1\linewidth]{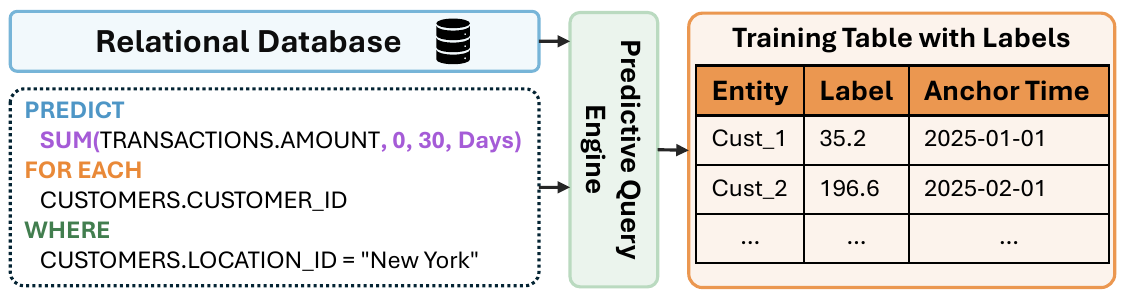}
    \caption{\textbf{Predictive Query Overview}. \textnormal{PQL allows for declaring a prediction task by formulating a query: The \textcolor{predict}{\textbf{PREDICT}} statement describes \emph{what} is to be predicted and the \textcolor{foreach}{\textbf{FOR EACH}} and \textcolor{where}{\textbf{WHERE}} statements for \emph{whom} the prediction is to be made. In this example, the query is specifying a regression task of predicting the sum of transaction amounts over the next 30 days for all customers coming from New York.
    The predictive query engine then parses the query and generates a training table with labels from the database. The entries can be used to train a machine learning model or as examples for in-context learning in relational foundation models.}}
    \label{fig:overview}
\end{figure}

Training table generation is a central yet notoriously difficult step in applying machine learning to relational data. Real-world databases are dynamic: entities evolve over time, new records are appended, and historical values are updated or corrected. Constructing a correct training table, therefore, requires time travel---the ability to reason about the state of the database as it existed at a specific prediction time—and strict point-in-time consistency, ensuring that every feature value used for prediction was available at the moment the prediction would have been made. 

Consider the following scenario: The goal is to predict how much each customer in the database will spend on products in the next 30 days. Training label computation for such a task requires identifying all customers, i.e. \emph{prediction entities}, and aggregating individual transactions of each customer over the future 30 day interval in historical data. Importantly, for each training example, input features from the database need to be filtered to lie before the start of the 30 day interval.
Although such label computation can be implemented in existing languages such as SQL or libraries such as Pandas \cite{reback2020pandas}, this process tends to be slow and error prone, given that these tools were not designed with label computation as their main goal. 

\begin{figure*}[t]
    \centering
    \includegraphics[width=1\linewidth]{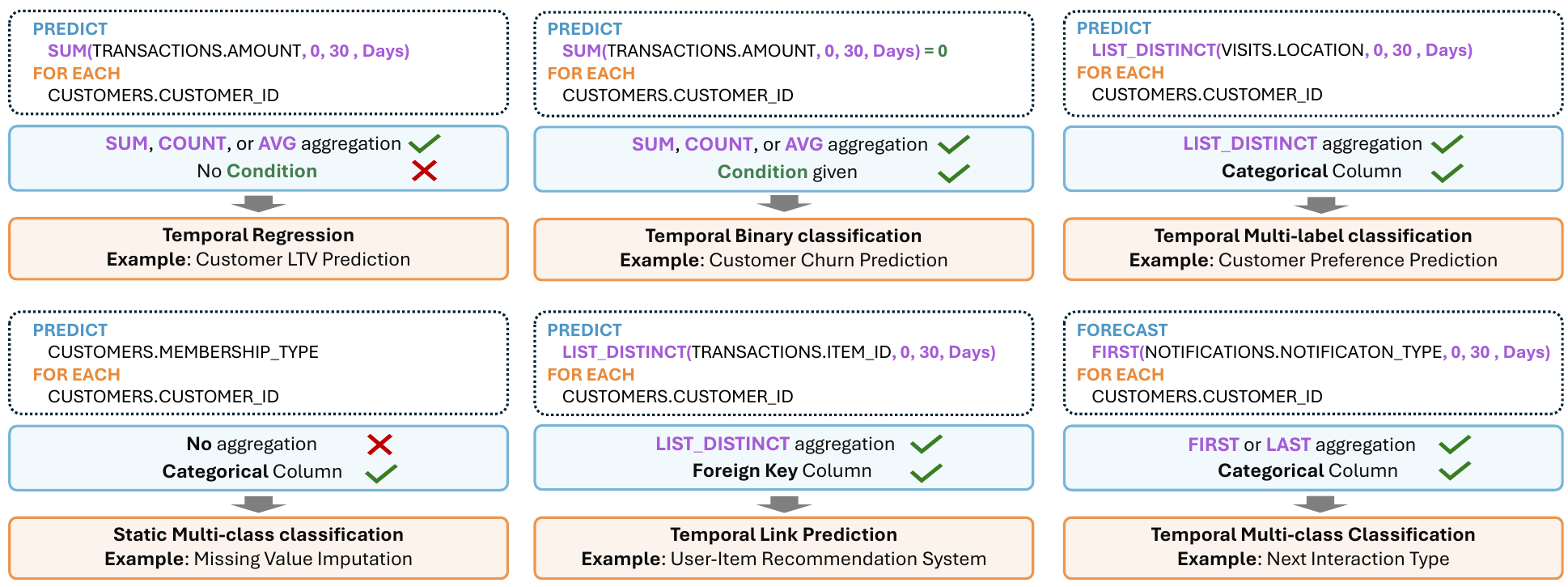}
    \caption{\textbf{Taxonomy of the Predictive Query Language (PQL)}. \textnormal{Examples of predictive queries written in the proposed PQL. The language allows declaration of a wide range of prediction tasks, such as regression, classification, forecasting, and link prediction. Typical applications can be LTV prediction or forecasting, missing value imputation, churn prediction, or item recommendation for customers.}}
    \label{fig:taxonomy}
\end{figure*}

Moreover, failing to enforce these constraints easily leads to label or information leakage, where future information (e.g., outcomes, post-event attributes, or aggregates computed over future data) inadvertently influences the model during training, resulting in overly optimistic performance estimates that do not generalize to deployment. These issues are subtle, ubiquitous, and hard to detect, especially when feature and label engineering involves complex joins, temporal filters, and aggregations across multiple tables. As a result, the generation of training tables is often slow, manual, and error-prone, requiring deep domain knowledge and careful auditing. 

An automated approach to training table generation that natively enforces temporal semantics and leakage-free data construction would not only reduce errors but also significantly speed up and improve the overall process of machine learning model development. A good declarative ML system that aims to address these shortcomings should, therefore, allow the user to concisely define the predictive task in a declarative way so that the training examples can be dynamically computed as needed.



Here we introduce a novel domain-specific language for declarative machine learning: the \emph{Predictive Query Language (PQL)} (Fig.~\ref{fig:overview}). PQL is a declarative language, inspired by SQL, for defining predictive tasks on relational databases. With PQL, a complete predictive task can be described in a single declarative query, which allows automatic generation of training labels for a wide range of machine learning problems, including regression, classification, time-series forecasting, and recommender systems.

\begin{figure}[h] 
  \raggedright
  \small
  \hspace*{15pt}
  \texttt{\textcolor{predict}{\textbf{PREDICT}} 
  \textcolor{aggr}{\textbf{COUNT}(}TRANSACTIONS.*\textcolor{aggr}{\textbf{, 0, 3, months)}}} \\ \hspace*{15pt}
    \texttt{\textcolor{foreach}{\textbf{FOR EACH}} ARTICLES.ARTICLE\_ID} \\ \hspace*{15pt}
    \texttt{\textcolor{where}{\textbf{WHERE}} ARTICLES.ARTICLE\_TYPE = "shirt"}
  \caption{Predictive Query Example. \textnormal{An instance of a query to predict demand for shirts over the next three months.}
  }
  \label{example_query}
\end{figure}

In PQL (Fig.~\ref{example_query}) we can specify the predictive task in a few lines of code while enabling automatic label computation without temporal data leakage or incorrectly categorized data. PQL supports a wide range of common machine learning tasks, such as regression, classification, or link prediction, depending on the defined target type (Fig.~\ref{fig:taxonomy}).
Moreover, it is designed to allow easy inference with the relevant metadata such as task type, prediction timeframe, and sources of valid negative examples.

To illustrate PQL, we provide a motivating example in Figure~\ref{example_query}, predicting the demand for shirts over the next 3 months from a database schema as given in Figure~\ref{Schema_fig}.
The PQL query defines two things: (1) for which \emph{entities} in the database should the prediction be made. For example, \texttt{\textcolor{foreach}{\textbf{FOR EACH}} ARTICLES.ARTICLE\_ID \textcolor{where}{\textbf{WHERE}} ARTICLES.ARTICLE\_TYPE = "shirt"} states that the prediction should be made for all articles whose type is a shirt.
(2) PQL defines what the prediction target is for each entity. For example, \texttt{\textcolor{predict}{\textbf{PREDICT}} \textcolor{aggr}{COUNT(}TRANSACTIONS.*\textcolor{aggr}{, 0, 3, MONTHS)}} defines the prediction target of predicting the count of purchases of each shirt by implicitly defining it as the count of all transactions over the next 3 months. From the above PQL we can also infer the task to be a regression problem and enable the automatic choice of an appropriate model.

Furthermore, given a defined set of valid entities and their corresponding targets, PQL can be used to generate training data following the Relational Deep Learning (RDL) paradigm~\cite{rdl}.
RDL paradigm generates training instances for an ML model by looking at earlier time windows in the database for which the labels can be computed. For this, \emph{anchor times} are sampled, historical points in time for which appropriate future training label intervals exist, i.e., the starting points of the 3 month intervals in Figure~\ref{example_query}~\cite{rdl}.
In the training process, one can then ensure that only the input features from before the anchor time are used to predict the corresponding target during training. We discuss details of this process in Section~\ref{section:training_table_construction}.

\begin{figure*}[t] 
  \centering
\footnotesize  

\begin{tikzpicture}[
  table/.style={
    rectangle split,
    rounded corners=.1cm,
    draw,
    text width=2.4cm,
    minimum height=1cm,
    align=left,
    font=\sffamily
  },
  arrow/.style={-{Latex[length=2mm]}, thick}
]

\node (notifications) [
  table,
  rectangle split parts=5,
  rectangle split part fill={
    gray!30,    
    white,      
    white,      
    white,      
    foreach!20      
  }
] {
  \textbf{Notifications}
  \nodepart{two} notification\_type\\ (string, categorical)
  \nodepart{three} notification\_text\\ (string, text)
  \nodepart{four} time\_sent \\(timestamp, time)
  \nodepart{five} customer\_id \\(FK)
};

\node (customers) [
  table,
  rectangle split parts=5,
  rectangle split part fill={
    gray!30,    
    predict!30,    
    white,      
    white,      
    white       
  },
  right=1.3cm of notifications
] {
  \textbf{Customers}
  \nodepart{two} \underline{customer\_id} \\(PK)
  \nodepart{three} location\_id\\ (str, categorical)
  \nodepart{four} signup\_date\\ (timestamp, temporal)
  \nodepart{five} membership\_type\\ (str, categorical)
};

\node (transactions) [
  table,
  rectangle split parts=6,
  rectangle split part fill={
    gray!30,    
    predict!30,    
    white,      
    white,      
    foreach!20,     
    foreach!20      
  },
  right=1.3cm of customers
] {
  \textbf{Transactions}
  \nodepart{two} \underline{transaction\_id} \\(PK)
  \nodepart{three} value \\(float, numerical)
  \nodepart{four} timestamp \\(timestamp, temporal)
  \nodepart{five} customer\_id \\(FK)
  \nodepart{six} article\_id \\(FK)
};

\node (articles) [
  table,
  rectangle split parts=6,
  rectangle split part fill={
    gray!30,    
    predict!30,    
    white,      
    white,      
    white,      
    white       
  },
  right=1.3cm of transactions
] {
  \textbf{Articles}
  \nodepart{two} \underline{article\_id} \\(PK)
  \nodepart{three} article\_name \\(string, text)
  \nodepart{four} article\_type \\(string, categorical)
  \nodepart{five} description \\(string, text)
  \nodepart{six} color \\(string, categorical)
};

\node (trainingtable) [
  table,
  rectangle split parts=4,
  rectangle split part fill={
    gray!30,    
    aggr!60,      
    aggr!20,      
    aggr!20      
  },
  right=1cm of articles
] {
  \textbf{Training Table}
  \nodepart{two} Entity\\ (FK)
  \nodepart{three} Target \\ (int, numerical)
  \nodepart{four} Timestamp \\(timestamp, time)
};

\draw[arrow] (transactions.five west) -- ++(-0.65,0) |- (customers.two east);
\draw[arrow] (transactions.six east) -- ++(0.65,0) |- (articles.two west);
\draw[arrow] (notifications.five east) -- ++(0.65,0) |- (customers.two west);
\draw[arrow, dotted] (trainingtable.two west) -- ++(-0.5,0) |- (articles.two east);

\end{tikzpicture}

  \caption{Example database schema. \textnormal{An example retail database consisting of all customers, articles, and their timestamped transactions and notifications, serving as basis for the running query examples. Each row contains information about the data type and semantic type, with links between keys marked with arrows. Primary keys (PK) are highlighted in \textcolor{predict}{blue} and foreign keys (FK) in \textcolor{foreach}{orange}. If PQL example from Figure~\ref{example_query} is used on this query, the training table (\textcolor{aggr}{violet}) is added to the database, with three columns: \texttt{Entity}, pointing to the entity of the query, \texttt{Target}, denoting the label---count of the transactions, and \texttt{Timestamp}, denoting the anchor time from which the label was computed.}}
  \label{Schema_fig}
\end{figure*}

To support a broad coverage of practical predictive problems, PQL supports a collection of features.
To define a valid set of entities, PQL supports filtering both on past and present data, e.g., only predict for customers in New York that have purchased an item in the last month, as well as making assumptions for testing counterfactuals, e.g., predicting purchases under the assumption that the customer received a notification in the next day. To define labels, as well as filters, PQL supports a rich set of aggregations and operations that allow both aggregation of values, e.g. \textcolor{aggr}{\textbf{\texttt{SUM}}} of purchased item values, or predicting the existence of new entries to the database, e.g. \textcolor{aggr}{\textbf{\texttt{LIST\_DISTINCT}}} over their keys.
When combined with an appropriate machine learning model, it creates an interactive system where creating new predictions about the future is as simple as querying the existing data.


We present two implementations of PQL.
The first is the implementation created for the Relational Deep Learning~\cite{rdl} framework and serves large-scale use cases in which models are trained for each query individually. The second implementation is designed for a Relational Foundation Model~\cite{kumorfm} to answer predictive queries within a second, requiring a low-latency, interactive implementation in which PQL is used to obtain examples for in-context learning.
We tested our PQL implementations on a collection of popular benchmarks for relational deep learning~\cite{relbench} and find that our custom implementation computes the training table up to $40$-times faster compared to the baseline implementation. 
We find that the language is sufficiently expressive to serve multiple industry use cases, including powering recommendations on popular social media platforms and food delivery services, fraud detection on the Bitcoin blockchain, and predicting potential medical complications for a clinic.

The rest of the paper is structured as follows: First, we discuss the related work and the flaws of the existing approaches to predictive modeling through declarative syntax. Then, we introduce the Predictive Query Language by listing the requirements, followed by the syntax and its core features. The final section is a description of the two practical implementations of the language, including a collection of optimizations specific to this problem.

\section{Related Work}
To the best of our knowledge, there is no existing work that attempts to solve the problem of training label generation for relational deep learning.
However, over the years, many systems have been introduced to simplify machine learning workflows through declarative languages and configurations.
In this section, we divide them and review them according to their type and compare them with PQL.

\xhdr{Configuration-Based Declarative ML}
Classical approaches to declarative ML focus on translating high-level pipeline or model definition into scalable pipelines, focusing on efficiently processing input features~\cite{xin2018helix, DeepFeatureSynthesis, lam2017one}, efficiently defining models~\cite{ludwig, overton}, or implementing ML algorithms at scale~\cite{systemml1, systemml2, sparks2017keystoneml, tfx}.
These systems usually introduce a configuration-based declarative language that defines an abstract pipeline optimized and materialized by the system.
Ludwig platform~\cite{ludwig}, for example, supports the definition of encoder-decoder models on multiple data modalities including text, audio, images and categorical data with a YAML-based approach that emphasizes ease of use and rapid prototyping.
{Overton}~\cite{overton}  takes a similar configuration-driven approach, but focuses on production deployment and monitoring for large-scale systems. It excels at multitask learning and operational concerns like A/B testing.
Unlike PQL, these systems assume that labels are readily available and focus on model or data pipeline configuration.
These only represent a part of the ML workflow and are increasingly sidetracked by advancements in foundation models~\cite{kumorfm} and AutoML that reduce the need for manual iteration and manually defined pipelines.

\xhdr{AutoML Platforms} Auto ML platforms and tools focus on a specific task within the ML pipeline: finding the best model for a well-defined task~\cite{mlbase,erickson2020autogluon,autosklearn, tpot,thornton2013autoweka}.
An example of such a system is {MLBase}~\cite{mlbase}, where users specify input features and target variables, and the system automatically selects learning algorithms and performs parameter tuning. MLBase demonstrated the value of automated machine learning for traditional tabular machine learning tasks, enabling users to focus on problem definition rather than model selection and hyperparameter optimization.
Using PQL and AutoML platforms is not mutually exclusive, as they ultimately tackle different problems; the latter can be used to train the model on the training table generated by the former.
However, we do not use any of the existing AutoML tools in our implementations described in Section~\ref{section:case_studies}.
The closest approach to PQL was the Feast tool~\cite{Kanter2016label}, however,due to the limitations of the underlying models, it still heavily focused on feature development and did not provide the required full expressivity.

\xhdr{Integrating ML into the Database} {ML2SQL}~\cite{ml2sql} translates machine learning code into database-specific stored procedures, enabling end-to-end workflows within PostgreSQL. The system provides a clear separation between the data preparation and model training phases but remains limited to traditional ML algorithms.
On the other hand, tools such as MAD skills~\cite{madskills} and {SQL4ML}~\cite{sql4ml} extend SQL to support machine learning operations directly within the database. This approach leverages existing SQL knowledge and infrastructure, which makes it particularly suitable for users with strong SQL expertise. 
These tools simplify integration of machine learning algorithms in a database environment, however, they ultimately do not reduce the ML knowledge required to train a model. 

\xhdr{Querying trained models} {GenSQL}~\cite{gensql}, ModelJoin~\cite{Klbe2023ExplorationOA}, and Raven~\cite{park2022raven} introduce a rich syntax for querying a probabilistic model, allowing multiple features such as probability estimation and data generation while seamlessly integrating with SQL.
However, these systems assume that a predictive model already exists and cannot be used to automatically create one.

These existing approaches share fundamental limitations when applied to deep learning on relational databases.
First, their design heavily focuses on features that are not required for relational deep learning, such as mapping the relational data into feature vectors, i.e.\ feature engineering or defining a model training pipeline.
Second, none provides intuitive abstractions for expressing entity-centric predictions that depend on complex aggregations across related tables or temporal windows. 
As a consequence, preventing temporal leakage through overlap and extracting metadata is left to the manual efforts of the user, making the process unnecessarily slow and prone to mistakes.
And last, converting relational databases to graph structures suitable for modern deep learning approaches like Graph Neural Networks~\cite{chen2025relgnn} or Graph/Relational Transformers~\cite{dwivedi2025relgt, ranjan2025relational}  requires substantial preprocessing and domain knowledge that these systems cannot provide. What is required is a declarative method that works natively with relational data, and our PQL is the first approach that meets this need.

\section{The Predictive Query Language}
This section introduces the Predictive Query Language (PQL) and its core features.
We define an entity- and timestamp-oriented training table and formalize the requirements such a language must satisfy.
Building on these foundations, we present the syntax of PQL and illustrate its capabilities, including automatic task inference, the incorporation of assumptions about the future, and support for both static and temporal queries.
Finally, we discuss how computing a single label can be repeated to construct the entire training and prediction table, along with key implementation details.

\subsection{Background: Training Tables}
\label{sec:training_table}

Relational Deep Learning (RDL)~\cite{rdl} describes a blueprint for machine learning directly on relational databases.
Given a training table whose examples consist of entities, anchor times, and target labels, RDL defines a procedure for automatically constructing and training a graph machine learning model.
The training table is linked to the rest of the database by connecting its entities to the corresponding entries in an entity table, which is itself related to other tables through primary--foreign key relationships.
From this structure, RDL automatically extracts subgraphs of connected table rows, serving as input features for the given training example. The whole blueprint eliminates the need for manual feature engineering and allows training graph or graph transformer models directly on raw database data. The benefits of RDL are: (1) the elimination of manual feature engineering, resulting in up to 20$\times$ faster model development, and (2) improved model precision from learning directly on raw data~\cite{relbench}.


More formally, a \emph{training table} in RDL is a collection of rows with the following columns (Figure~\ref{Schema_fig}): 
\begin{itemize}
    \item \emph{Entity}: A primary key that defines the entity from the database for which prediction is made. In RDL, it links individual training examples to their corresponding subgraphs of input features. Example: A valid primary key from the customer table.
    \item \emph{Target}: The target label which a machine learning model is trained to predict. Unlike the entity, the target can take any value or type, including lists when predicting multiple targets per entity---common for multilabel classification or recommendation. The only requirement is that all values in the column have the same type.
    \item \emph{Timestamp} (optional): For \emph{temporal} tasks, the training table includes an \emph{anchor time}. For example, if the aim is to predict transactions of a user over the next $30$ days, the \emph{anchor time} denotes the beginning of this window and enables filtering of input features inserted after it.
\end{itemize}

These columns constitute the minimal requirements for a well-defined specification of inputs and outputs, though additional columns may be included depending on the implementation, such as example weights or auxiliary features.

In Figure~\ref{fig:overview} we provide an example of a training table for an example query that aims to predict the spending patterns of New York based customers.
Each entity or anchor time may appear multiple times; however, each pair of (entity, anchor time) is unique, defining its own corresponding label. During model training, this training table is connected to the \texttt{customers} table via the \texttt{customer\_id} primary key, after which we can apply Relational Deep Learning to train a purchase value predictor.


\subsection{Requirements and Assumptions}
\xhdr{Language Requirements}
The following requirements are not guaranteed by existing data-processing languages such as SQL:
\begin{itemize}
    \item \textbf{Entity Definition:} The language must clearly define the set of valid entities, allowing the machine learning algorithm to dynamically select the relevant subset for training and prediction.
    \item \textbf{Target Definition:} The target should be a command that computes exactly one label per selected entity independently of other entities and targets. If the target is defined relative to a timestamp (e.g.\ ``count of transactions in the next week''), the target defines exactly one label per (entity, timestamp) pair. This is necessary to compute the labels for arbitrary subsets of entities.
    \item \textbf{Data Leakage Prevention:} Given an entity and a target, the language must allow identification of database records that could cause data leakage if used as input. This may occur either because these data are involved in computing the label or because they were inserted into the database after the \emph{anchor timestamp} associated with the input.
    \item \textbf{Inferrable Timeframe:} If the label is computed relative to some time, e.g. ``purchases in the next 30 days'', the exact timeframe needs to be inferrable from the query. This is essential for training data generation automation, as it allows generating labels from past examples without the risk of data leakage through temporal overlap.
    \item \textbf{Task Definition:} If the language supports multiple problem types, the task type must be inferrable from the query. Following~\citet{relbench}, we aim to support classification (binary and multi-class), regression, and link prediction (recommendations). On top of these, we identify the multi-label classification task type.
\end{itemize}

\xhdr{Assumptions on the Database Schema}
\label{sec:assumptions}
Following the RDL framework~\cite{rdl}, we assume that the database schema is known in advance and has clearly defined keys and timestamp columns (Figure~\ref{Schema_fig}).
Specifically, every key is either a \textbf{primary key} (a unique identifier for a database row) or a \textbf{foreign key} (a reference to a primary key)~\cite{database_systems}.
Tables that record time-dependent facts, such as events, contain a \textbf{time column} specifying when the event occurred.

Moreover, we assume that all entities are primary keys from a single column.
This may not always hold, for example, if we are making a prediction for every pair of items to predict co-occurrence.
However, this requirement can always be met through pre-processing and introducing the list of valid entities as an auxiliary table. In the example case, this would mean adding a table of combinations of pairs of items of interest.

Finally, we assume that each column is annotated with a data type (\textit{int, bool, string, \ldots}) and a semantic type (\textit{numerical, categorical, text, \ldots}), following the types defined in the PyTorch Frame framework~\cite{hu2024pytorch}.
These can be either auto-inferred or user-specified.

We find these assumptions to be non-restrictive in practice, as they align with standard database design principles and are already required by the Relational Deep Learning methods for which labels are generated.

\subsection{PQL Syntax}
We define the following syntax for our Predictive Query Language (PQL), which meets the requirements outlined above.
PQL decomposes the specification of a predictive problem on relational data into two components:
\begin{enumerate}
\item \textcolor{predict}{\textbf{\texttt{PREDICT}}}: Specifies the prediction target, computed for each entity.
\item  \textcolor{foreach}{\textbf{\texttt{FOR EACH}}}: Specifies the set of entities for which predictions are made, along with any relevant filters.
\end{enumerate}
\begin{figure}[t] 
  \raggedright
  \small
  \hspace*{15pt}
  \textcolor{predict}{\texttt{\textbf{PREDICT}}} \texttt{TRANSACTIONS.VALUE} \\ \hspace*{15pt}
  \textcolor{foreach}{\textbf{\texttt{FOR EACH}}} \texttt{TRANSACTIONS.TRANSACTION\_ID}
  \caption{Example of a static predictive query. \textnormal{The query defines each \texttt{TRANSCTION\_ID} to be an entity, and its corresponding \texttt{VALUE} to be the target. In this  query there is always exactly one target per entity.}
  }
  \label{simple_static_query}
\end{figure}

\begin{figure}[t]
    \small
    \begin{verbatim}
grammar PQLGrammar;

prog:
    PREDICT target problem_type top_k
    'FOR EACH' entity assuming EOF;

// Query sections
target: condition | aggregation | column;
entity: column | filtered_column;
assuming: ('ASSUMING' condition)?;
problem_type: ('RANK' | 'CLASSIFY')?;
top_k: ('TOP' INT)?;

// operations
condition:
    aggregation REL_OP constant
    | column REL_OP constant
    | 'NOT' condition
    | condition 'AND' condition
    | condition 'OR' condition
    | '(' condition ')';
    
aggregation:
    AGGR_TYPE '(' (column | filtered_column) ',' 
               (INT | '-INF') ',' INT ')'
    | AGGR_TYPE '(' (column | filtered_column) ',' 
               (INT | '-INF') ',' INT ',' TIME_UNIT ')'
    | AGGR_TYPE '(' (column | filtered_column) ')';
    
filtered_column: column WHERE condition;
    
column: NAMED_COLUMN | WILDCARD_COLUMN;
    
constant: 
    BOOL | INT | STR | DECIMAL | datetime | NULL | array;
    
NAMED_COLUMN: ID '.' ID;
WILDCARD_COLUMN: ID '.' '*';
    
AGGR_TYPE: 
    'SUM' | 'AVG' | 'MIN' | 'MAX' | 'COUNT_DISTINCT' 
    | 'FIRST' | 'LAST' | 'LIST_DISTINCT' | 'COUNT';

REL_OP: 
    '!=' | '<=' | '>=' | '<' | '>' | '=' | 'IS' 
    | 'IS NOT' | 'IN' | 'IS IN' | 'LIKE' | 'NOT LIKE'
    | 'CONTAINS' | 'NOT CONTAINS' 
    | 'STARTS WITH' | 'ENDS WITH';
    \end{verbatim}
    \vspace{-5mm}
    \caption{PQL grammar. \textnormal{The grammar is implemented in ANTLR4 syntax~\cite{antlr4}. Parts of the grammar, including selected terminal symbols, are omitted for brevity.}}
    \label{PQLGrammar}
\end{figure}
Figure~\ref{simple_static_query} shows an example of a simple query that predicts missing transaction values in the database, using all available transaction values as training data. The full grammar of PQL is provided in Figure~\ref{PQLGrammar}, with non-essential components omitted for brevity.

PQL supports three core operations: aggregations, conditions, and filters. An \textcolor{aggr}{\emph{aggregation}} aggregates values of a single column within a certain time window, grouped by a foreign key.
For example,\\ \hspace*{15pt}
\texttt{\textcolor{predict}{\textbf{PREDICT}} \textcolor{aggr}{SUM(}TRANSACTIONS.VALUE\textcolor{aggr}{, 0, 30, days)}\\ \hspace*{12.7pt}
\textcolor{foreach}{\textbf{FOR EACH}} CUSTOMERS.CUSTOMER\_ID} \\
predicts a sum of all transaction values $30$ days starting from the anchor timestamp, grouped by \texttt{CUSTOMER\_ID}.

A \textcolor{where}{\emph{condition}} is an operation with a boolean value, either a logical operation, or a comparison of an aggregation or a column with a value.
Finally, a \emph{filter}, expressed with the keyword \textcolor{where}{\texttt{WHERE}}, is a mechanism that selects a subset of the data that meets a condition.
For example, counting the number of transactions above $\$100$ for all customers with at least one transaction in the past $30$ days:
\\ \hspace*{15pt}
\texttt{\textcolor{predict}{\textbf{PREDICT}} \textcolor{aggr}{COUNT(}\\ \hspace*{30pt}
TRANSACTIONS.* \textcolor{where}{\textbf{WHERE}} TRANSACTIONS.VALUE \textcolor{where}{> 100}, \\ \hspace*{30pt}
\textcolor{aggr}{0, 30, days} \\ \hspace*{13pt}
\textcolor{aggr}{)}\\ \hspace*{13pt}
\textcolor{foreach}{\textbf{FOR EACH}} CUSTOMERS.CUSTOMER\_ID\\ \hspace*{13pt}
\textcolor{where}{\textbf{WHERE}} \textcolor{aggr}{COUNT(}TRANSACTIONS.*\textcolor{aggr}{, -30, 0, days)} \textcolor{where}{> 0}
} \\

The following sections explain how these elements of the PQL grammar can be combined to form semantically meaningful queries.

\subsection{Handling Static Data}

The query in Figure~\ref{complex_static_query}
is an instance of a \emph{static} query---a query without temporal aggregations where each entity is associated with exactly one label.
Naturally, predicting values that exist in the database has limited practical value.
We thus assume that the query is used when \texttt{TRANSACTION.VALUE} is missing for at least some rows and the machine learning model will be used to impute those values.
Note that filtering transactions by customer location is sound and supported because exactly one foreign key points to a single row in the \texttt{CUSTOMERS} table.
The schema therefore guarantees that the \textcolor{where}{\texttt{WHERE}} filter is well-defined.

\subsection{Handling Temporal Data}
Most predictive tasks are not about filling in the gaps in the database.
Instead, we usually want to predict behavior of the database in the future, e.g.\ predicting future purchases.
To condense future entries of a database into a single label, PQL utilizes aggregations, e.g.\ predicting the \textcolor{aggr}{\texttt{SUM}} of all future purchases or generating a list of all purchased articles (aggregation \textcolor{aggr}{\texttt{LIST\_DISTINCT}}).
An example of a temporal query with aggregations is given in Figure~\ref{temporal_query}.

\begin{figure}[t] 
  \raggedright
  \small
  \hspace*{15pt}
  \texttt{\textcolor{predict}{\textbf{PREDICT}} TRANSACTIONS.VALUE \textcolor{where}{> 100} \\ \hspace*{13pt}
  \textcolor{foreach}{\textbf{FOR EACH}} TRANSACTIONS.TRANSACTION\_ID \\ \hspace*{13pt}
  \textcolor{where}{\textbf{WHERE}} CUSTOMERS.LOCATION\_ID \textcolor{where}{= "New York"}}
  \caption{Static query. \textnormal{Predictive query that predicts whether missing transaction values of New York customers are over \$100.}}
  \label{complex_static_query}
\end{figure}

For the aggregation to be well defined, it must operate on a timestamp-annotated target table that is directly connected to the entity table with a foreign key.
In Figure~\ref{Schema_fig}, \texttt{the transactions} contain a foreign key to the table \texttt{articles}, implying that there may be multiple transactions per article.
Restricting the use of aggregations to tables with semantically meaningful connections in the schema is not just a matter of syntactic convenience, but rather has a practical purpose.
The schema of the database is used by machine learning models to make predictions.
If a connection between keys is relevant to the label computation, it most likely carries valuable semantic information that is relevant to the model and should therefore be explicitly included in the schema.

Note that aggregations for an individual entity are always computed relative to the anchor time.
The syntax of PQL unambiguously defines the timeframe (85 days in Figure~\ref{temporal_query}), allowing the selection of anchor times that do not leak information through overlap.

The exact choice of entities and anchor timestamps in the context data is specific to the use case, i.e.\ how much data the predictive model requires.
When generating future-looking predictions, the value of the prediction anchor time can be set to the timestamp of interest, e.g.\ current time.

\subsection{Inferring the Task Type}

Given a query, PQL implicitly defines its task type.
Based on the data and semantic types of the referenced columns as well as the query structure, the data and semantic type of the computed labels can be determined unambiguously.
The decision process is indicated in Figure~\ref{fig:taxonomy} and depends on the target type.
Automatic task inference is a deliberate language design choice that enables automated model selection and task-specific handling.

We give special consideration to the \textcolor{aggr}{\texttt{LIST\_DISTINCT}} aggregation.
Unlike most other aggregations whose output is a single value, \textcolor{aggr}{\texttt{LIST\_DISTINCT}} returns a set of values, leading to a multi-label classification or a link prediction task; the latter happens when the aggregated column is a foreign key.
This case requires special handling; the goal is not to predict a single value, but an existence of a future entry into the database, i.e., a future transaction.
Training a link prediction model requires additional metadata that can be extracted from the query and database schema, such as negative samples and valid targets that have not appeared in the training data, e.g., new articles.

\begin{figure}[t] 
  \centering
\begin{tcolorbox}[colback=white, colframe=black, rounded corners, boxrule=0.5pt, width=8.5cm, boxsep=0pt]
\begin{minipage}{\linewidth}
\texttt{
\textcolor{predict}{PREDICT} \\ \hspace*{10pt}
\colorbox{lightblue}{\textcolor{aggr}{SUM(}transactions.value\textcolor{aggr}{,\ 15,\ 45, days)}} \textcolor{where}{$>$ 100} \\ \hspace*{10pt} OR \\ 
\hspace*{10pt} \colorbox{lightred}{\textcolor{aggr}{COUNT(}transactions.*\textcolor{aggr}{,\,15,\,45, days)}} \textcolor{where}{$>$ 10}\\
\textcolor{foreach}{FOR EACH} customers.customer\_id \\
\hspace*{10pt} \textcolor{where}{WHERE} \colorbox{lightyellow}{\textcolor{aggr}{COUNT(}transactions.*\textcolor{aggr}{,\,{-}40,\,0, days)}} \textcolor{where}{$>$ 0}
}
\end{minipage}
\end{tcolorbox}

\begin{center}
\begin{tikzpicture}[x=0.8cm, y=0.8cm]

\draw[dotted, thick] (5,0.4) -- (5,-1.7);
\draw[dotted, thick] (10,0.4) -- (10,-1.7);

\draw[<->] (5.2,-1.5) -- (9.8,-1.5) node[midway, below] {\footnotesize 45 days};
\node[anchor=north] at (5,-1.9) {\textbf{\footnotesize Anchor Time}};

\node[anchor=south west] at (5.95,0.35) {\scriptsize \textcolor{aggr}{SUM(}transaction.value\textcolor{aggr}{,\ 15,\ 45, days)}};
\draw[fill=lightblue] (6.6,0.45) rectangle (10,0.20);

\node[anchor=south west] at (6.02,-0.30) {\scriptsize \textcolor{aggr}{COUNT(}transactions.*\textcolor{aggr}{,\,15,\ 45, days)}};
\draw[fill=lightred] (6.6,-0.2) rectangle (10,-0.45);

\node[anchor=south west] at (0.5,-0.8) {\scriptsize \textcolor{aggr}{COUNT(}transactions.*\textcolor{aggr}{,\,{-}40,\ 0, days)}};
\draw[fill=lightyellow] (0.5,-0.7) rectangle (5,-0.95);

\end{tikzpicture}
\end{center}
  \caption{Timeline example. \textnormal{Example of a temporal predictive query that predicts whether an active user will either spend at least 100 dollars in the time period between 15 and 45 days in the future, or have at least 10 transactions. The image below illustrates the decomposition of a single timeframe that operates on 85 days of data (40 days in the past and 45 in the future of the anchor timestamp).}}
  \label{temporal_query}
\end{figure}



\subsection{Conditioning on the Future}

The example in Figure~\ref{assuming_query} additionally contains an \texttt{ASSUMING} clause, a future-looking filter that describes an assumption about the training data.
This adds an option to express counterfactual conditions for causal inference~\cite{appliedcausal}.
Intuitively, it operates as a forward-looking filter applied during label-generation time.

By operating as a forward-looking filter, \texttt{ASSUMING} creates a training distribution of examples that meet the assumption.
In the example in Figure~\ref{assuming_query}, the assuming operator restricts the training examples to cases where the user received a notification.
A machine learning model trained on such a training table will be conditioned to generate its predictions under the assumption that the notifications were sent, allowing decision making based on hypothetical scenarios.
By comparing the predictions of a model obtained with the query in Figure~\ref{assuming_query} with the query in Figure~\ref{temporal_query}, one can estimate the impact of sending a notification.

Despite acting as a filter on the training table, this operator requires separate syntax.
This is not only due to its semantic role but also to allow its removal when generating the entity list for future predictions.


\begin{figure}[t] 
  \centering
\begin{tcolorbox}[colback=white, colframe=black, rounded corners, boxrule=0.5pt, width=8.5cm, boxsep=0pt]
\begin{minipage}{\linewidth}
\texttt{
\textcolor{predict}{PREDICT} \\ \hspace*{10pt}
\colorbox{lightblue}{\textcolor{aggr}{SUM(}transactions.value\textcolor{aggr}{,\ 15,\ 45, days)}} \textcolor{where}{$>$ 100}\\ \hspace*{10pt} OR \\ 
\hspace*{10pt} \colorbox{lightred}{\textcolor{aggr}{COUNT(}transactions.*\textcolor{aggr}{,\,15,\,45, days)}} \textcolor{where}{$>$ 10}\\
\textcolor{foreach}{FOR EACH} customers.customer\_id \\
\hspace*{10pt} \textcolor{where}{WHERE} \colorbox{lightyellow}{\textcolor{aggr}{COUNT(}transactions.*\textcolor{aggr}{,\,{-}40,\,0, days)}} \textcolor{where}{$>$ 0}\\
\textbf{ASSUMING} \\ \hspace*{10pt}
\colorbox{lightgreen}{\textcolor{aggr}{COUNT(}notifications.*\textcolor{aggr}{,\,0,\,15, days)}} \textcolor{where}{$>$ 0}
}
\end{minipage}
\end{tcolorbox}

\vspace{-0.2cm}

\begin{center}
\begin{tikzpicture}[x=0.8cm, y=0.8cm]

\draw[dotted, thick] (5,0.4) -- (5,-2.2);
\draw[dotted, thick] (10,0.4) -- (10,-2.2);

\draw[<->] (5.2,-2.0) -- (9.8,-2.0) node[midway, below] {\footnotesize 45 days};
\node[anchor=north] at (5,-2.4) {\textbf{\footnotesize Anchor Time}};

\node[anchor=south west] at (5.95,0.35) {\scriptsize \textcolor{aggr}{SUM(}transaction.value\textcolor{aggr}{,\ 15,\ 45, days)}};
\draw[fill=lightblue] (6.6,0.45) rectangle (10,0.20);

\node[anchor=south west] at (6.03,-0.30) {\scriptsize \textcolor{aggr}{COUNT(}transactions.*\textcolor{aggr}{,\,15,\ 45, days)}};
\draw[fill=lightred] (6.6,-0.2) rectangle (10,-0.45);

\node[anchor=south west] at (0.5,-0.8) {\scriptsize \textcolor{aggr}{COUNT(}transactions.*\textcolor{aggr}{,\,{-}40,\ 0, days)}};
\draw[fill=lightyellow] (0.5,-0.7) rectangle (5,-0.95);

\node[anchor=south west] at (5.05,-1.5) {\scriptsize \textcolor{aggr}{COUNT(}notifications.*\textcolor{aggr}{,\,0,\ 15, days)}};
\draw[fill=lightgreen] (5,-1.4) rectangle (6.6,-1.65);

\end{tikzpicture}
\end{center}
  \caption{Timeline example. \textnormal{In this example we augment the example from Figure~\ref{temporal_query} with an assumption that the customer is sent a notification in the next 15 days. The \texttt{ASSUMING} filter is only applied during training, despite being future-looking.}
  }
  \label{assuming_query}
\end{figure}

\subsection{Training Table Construction}
\label{section:training_table_construction}
The preceding sections describe how to define the label for an individual entity or an (entity, timestamp) pair.
Constructing the full training table, as defined in Section~\ref{sec:training_table} takes repeating this procedure for a suitable set of examples.
While the exact details of entity and timestamp selection are implementation-specific and can be found in Section~\ref{section:case_studies}, certain points are common across implementations.

For static queries, labels are computed for all entities in the entity table.
Wherever the resulting label cannot be computed due to missing data, the example is ignored during model training.
The remaining examples are split according to the training, validation, and test splits.
The examples that were dropped during training due to missing labels are used as inputs during prediction time, i.e.\ we predict the values for entities where it could not be computed.

If the query is temporal, the timestamps corresponding to the anchor times can be selected gradually, moving back in time, as demonstrated in Figure~\ref{anchor_sampling}.
To prevent data leakage, the split into training, validation, and test sets is typically performed at the timestamp level, ensuring that all examples sharing the same timestamp belong to the same split.
Unlike in the static case, there is no data held out for prediction as predictions target future entity behavior.
To avoid any temporal overlap between examples in different sets that could lead to accidental information leakage, the timestamps are usually selected one timeframe apart.
In this way, the data separation boundaries can be safely selected at any timestamp.
The label is then generated for every valid pair of entities and selected timestamps.
Because the number of selected timestamps directly determines the size of the training table and the computational cost, it is highly implementation- and use case–dependent, as discussed in Section~\ref{section:case_studies}.

During prediction time, the \textcolor{predict}{\texttt{\textbf{PREDICT}}} and \texttt{\textbf{ASSUMING}} clauses are ignored, as labels are predicted rather than computed.
The exception is the link prediction setting, such as the one given in Figure~\ref{complex_target_filter} --- here, the target table of the foreign key in the \textcolor{aggr}{\textbf{\texttt{LIST\_DISTINCT}}} aggregation is used as the source of valid targets.

\begin{figure}[t] 
  \centering
\begin{tcolorbox}[colback=white, colframe=black, rounded corners, boxrule=0.5pt, width=8cm]
\begin{minipage}{\linewidth}
\texttt{
\textcolor{predict}{PREDICT} \\ \hspace*{10pt}
\colorbox{lightblue}{\textcolor{aggr}{SUM(}transactions.value\textcolor{aggr}{,\ 0,\ 30, days)}} \\
\textcolor{foreach}{FOR EACH} customers.customer\_id 
}
\end{minipage}
\end{tcolorbox}

\vspace{-0.2cm}

\begin{center}
\begin{tikzpicture}[x=0.8cm, y=0.8cm]

\draw[dotted, thick] (8.5,0.4) -- (8.5,-3.5);

\draw[<->] (8.55,-3.1) -- (10,-3.1) node[midway, below] {\footnotesize Future};
\draw[<->] (0,-3.1) -- (8.45,-3.1) node[midway, below] {\footnotesize Existing Historical Data};
\node[anchor=north] at (8.5,-3.4) {\textbf{\footnotesize Current Time}};

\node[anchor=south west] at (4.3,0.45) {\footnotesize \textbf{Prediction}};
\draw[fill=lightblue] (8.5,0.4) rectangle (10,0.1);
\draw[fill=gray!40] (0.0,0.4) rectangle (8.5,0.1);
\node[anchor=south west] at (5.3,-0.03) {\scriptsize $\leftarrow$ Potential Input Features };
\node[anchor=south west] at (8.8,-0.03) {\scriptsize Target};
\draw[ultra thick] (8.5,0.5) -- (8.5,0);

\draw[<->] (8.55,-0.1) -- (10,-0.1) node[midway, below] {\footnotesize 30 days};

\node[anchor=south west] at (3.05,-0.60) {\footnotesize \textbf{Training or Context Examples}};

\draw[fill=lightblue] (7,-0.6) rectangle (8.5,-0.9);
\draw[fill=gray!40] (0.0,-0.6) rectangle (7,-0.9);
\draw[ultra thick] (7,-0.5) -- (7,-1);
\node[anchor=south west] at (6,-1.4) {\footnotesize Anchor time};
\node[anchor=south west] at (3.8,-1.03) {\scriptsize $\leftarrow$ Potential Input Features };
\node[anchor=south west] at (7.3,-1.03) {\scriptsize Target};

\draw[fill=lightblue] (5.5,-1.4) rectangle (7,-1.7);
\draw[fill=gray!40] (0.0,-1.4) rectangle (5.5,-1.7);
\draw[ultra thick] (5.5,-1.3) -- (5.5,-1.8);
\node[anchor=south west] at (4.5,-2.2) {\footnotesize Anchor Time};
\node[anchor=south west] at (2.3,-1.83) {\scriptsize $\leftarrow$ Potential Input Features };
\node[anchor=south west] at (5.8,-1.83) {\scriptsize Target};

\draw[fill=lightblue] (4,-2.2) rectangle (5.5,-2.5);
\draw[fill=gray!40] (0.0,-2.2) rectangle (4,-2.5);
\draw[ultra thick] (4,-2.1) -- (4,-2.6);
\node[anchor=south west] at (3,-3) {\footnotesize Anchor Time};
\node[anchor=south west] at (0.8,-2.63) {\scriptsize $\leftarrow$ Potential Input Features };
\node[anchor=south west] at (4.3,-2.63) {\scriptsize Target};

\end{tikzpicture}
\end{center}
  \caption{Anchor time sampling. \textnormal{Anchor times can be sampled from historical data for each entity. Each sampled anchor time defines a training/validation/context example, where the target is obtained from after the anchor time and input features are obtained from the time before, according to the RDL blueprint~\cite{rdl}. Different temporal sampling strategies can be used, such as random times, equal distances or oversampling more recent anchor times. }}
  \label{anchor_sampling}
\end{figure}


\subsection{Implementation}
\label{section:implementation}
The PQL parsing and validation follow the standard practices for compiler implementation~\cite{compilers}.
 Grammar and its parsing are implemented using the ANTLR4~\cite{antlr4} library.
The obtained syntax tree is transformed into an abstract syntax tree, which we use to implement multiple validations and ensure consistency with the database schema.
The validated abstract syntax tree is then transformed into a logical plan.
The logical plan serves as an intermediate representation where each node defines an operation to perform on a table in a data platform-agnostic manner.
In the process of these transformations, we define the order of operations, data access, internal variables, etc. following best practices for interpreter implementation~\cite{compilers}.

However, predictive task modeling differs from a regular querying language in a few ways. It therefore requires the following additional transformations, not included in a regular programming language interpreter:
\begin{itemize}
    \item \textbf{Join Inferral}: Based on the database schema, we infer the joins and keys used for each aggregation.
    \item \textbf{Mismatched Timestamp Filtering}: Suppose that entities in a temporal task come with a start date and end date, e.g.\ user sign-up and churn date. We add entity filters that drop any examples with a timestamp outside of the entity validity period.
    \item \textbf{Target Filter Separation for Prediction}: While label computation is usually ignored during prediction time, as there are no labels to generate, this is not the case for recommendations. In Figure~\ref{complex_target_filter}, the filter within \textcolor{aggr}{\texttt{LIST\_DISTINCT}} conditions on both transaction value and the article color. The condition \texttt{ARTICLES.COLOR = "blue"} can be extracted and applied at prediction time to score and recommend only articles of blue color.
\end{itemize}

\begin{figure}[t] 
  \raggedright
  \small
  \hspace*{15pt}
  \texttt{\textcolor{predict}{\textbf{PREDICT}} \textcolor{aggr}{LIST\_DISTINCT(} \\ \hspace*{30pt}
      TRANSACTIONS.ARTICLE\_ID \\ \hspace*{45pt}
      \textcolor{where}{\textbf{WHERE}} TRANSACTIONS.VALUE > 50 \\ \hspace*{45pt}
      \textbf{AND} ARTICLES.COLOR = "blue"\textcolor{aggr}{, \\ \hspace*{30pt}
      0, 30, days \\ \hspace*{15pt}
  )} \\ \hspace*{15pt}
  \textcolor{foreach}{\textbf{FOR EACH}} CUSTOMERS.CUSTOMER\_ID}
  \caption{Recommendation example. \textnormal{An instance of a recommendation query that only looks at transactions with value at least \$50 and articles with blue color. The label definition is still relevant at prediction time.
  The condition \texttt{ARTICLES.COLOR = "blue"} is extracted and applied to the list of valid targets (table \texttt{ARTICLES}).}}
  \label{complex_target_filter}
\end{figure}

Finally, the obtained logical plan undergoes implementation-specific optimization steps, which are described in further detail in Section~\ref{section:case_studies}.
The optimized logical plan can usually be directly translated into operations supported by most data processing libraries and engines, allowing straightforward scaling to multiple backends.


\section{Implementation and Case Studies}
\label{section:case_studies}
As a general specification framework for machine learning over both static and temporal relational data, PQL has been successfully used to generate training labels across a wide range of ML use cases. 
In this section, we introduce several real-world usecases and their corresponding queries, used in a production setting (with schemas anonymized for confidentiality).
We then describe two implementations and discuss practical aspects of their execution.

\subsection{Examples of Existing Usecases}

Figures~\ref{fig:taxonomy},~\ref{production_query_1},~\ref{production_query_2}, and~\ref{production_query_3} show example PQL queries inspired by actual queries used in the production setting.
Some of these have been successfully applied to databases with over 10 billion entries.

\begin{figure}[t] 
  \raggedright
  \small
  \hspace*{15pt}
  \texttt{\textcolor{predict}{\textbf{PREDICT}} \textcolor{aggr}{LIST\_DISTINCT(}\\ \hspace*{30pt}
  ORDERS.STORE\_ID\textcolor{aggr}{,\; 0,\; 7,\; days \\ \hspace*{13pt}
  ) RANK TOP 12}\\ \hspace*{13pt}
  \textcolor{foreach}{\textbf{FOR EACH}} USERS.CUSTOMER\_ID \\ \hspace*{13pt}
  \textcolor{where}{\textbf{WHERE}} USERS.LOC = 'NYC'}
  \caption{Example of a commercial recommendation usecase. \textnormal{Predict which store a customer in New York will order from in the next 7 days, used by a delivery platform.}}
  \label{production_query_1}
\end{figure}

\begin{figure}[t] 
  \raggedright 
  \small
  \hspace*{15pt}
  \texttt{\textcolor{predict}{\textbf{PREDICT}} \\ \hspace*{25pt}
  \textcolor{aggr}{COUNT(}PURCHASES.*\textcolor{aggr}{, 1, 4, days)} \textcolor{where}{> 0} \\ \hspace*{13pt}
  \textcolor{foreach}{\textbf{FOR EACH}} USERS.USER\_ID \\ \hspace*{13pt}
  \textcolor{where}{\textbf{WHERE}} \\ \hspace*{25pt}
  USERS.NOTIFICATION\_ELIGIBLE = 1 \\ \hspace*{13pt}
  \textbf{ASSUMING} \textcolor{aggr}{COUNT(} \\ \hspace*{25pt}
  NOTIFICATIONS.* \textcolor{where}{\textbf{WHERE}}
  NOTIFICATIONS.TYPE = 'PUSH'\textcolor{aggr}{,\\ \hspace*{25pt}
  0, 1, days \\ \hspace*{13pt}
  )} > 0}
  \caption{Example of a commercial usecase of counterfactual analysis. \textnormal{Predict whether a customer will make a purchase within 3 days of receiving a push notification. The example illustrates the use of the \texttt{ASSUMING} operator, conditioning on future events when generating labels useful for capturing causal or post treatment effects. 
  }}
  \label{production_query_2}
\end{figure}

\begin{figure}[t] 
  \raggedright
  \hspace*{15pt}
  \small
  \texttt{\textcolor{predict}{\textbf{PREDICT}} \textcolor{aggr}{LIST\_DISTINCT(}\\ \hspace*{30pt}
  USER\_PAGE\_VISITS.PAGE\_PATH\\ \hspace*{13pt}
  \textcolor{aggr}{) RANK TOP 10} \\ \hspace*{13pt}
  \textcolor{foreach}{\textbf{FOR EACH}} USERS.USER\_ID}
  \caption{Example of a commercial usecase of behavior modeling. \textnormal{Predict the top 10 most frequently visited URL paths for each user. This is a static query predicting which pages is a user most likely to visit, without relying on temporal slicing. This query is used as part of an advertising platform that models user behavior.}}
  \label{production_query_3}
\end{figure}

These examples showcase the flexibility of PQL in handling prediction tasks that involve complex temporal dependencies and user-defined filters. It has been used to deploy models in domains such as:

\begin{itemize}
    \item \textbf{Recommendation:} Predicting the top restaurants for users on a major food delivery platform.
    \item \textbf{Fraud Detection:} Identifying suspicious transaction patterns for a large financial institution.
    \item \textbf{Customer Lifetime Value (LTV):} Estimating the LTV for users of various online retail platforms.
\end{itemize}

In most of these cases, the database scale exceeds billions of rows and terabytes of data, requiring a robust and scalable implementation.

\subsection{Implementation: Batch Processing for Relational Deep Learning}


Relational Deep Learning~\cite{rdl} is a powerful paradigm for prediction on relational data enabled by recent advancements in graph neural networks.
Its defining feature is the elimination of manual feature engineering.
PQL plays a critical role in enabling RDL by generating training and prediction tables in a declarative and reproducible manner.

This implementation is primarily driven by scalability requirements. Enterprise databases often contain terabytes of data with billions of rows, and training high-quality models requires leveraging as many labels as possible. As a result, scalability and robustness are prioritized over low-latency execution.

To compute labels at scale, the logical plan obtained in Section~\ref{section:implementation} is translated into a physical Spark plan~\cite{spark} and batch-computed for every valid combination of entity and anchor timestamp.
Anchor timestamps are usually taken one timeframe apart to make use of the full data history while avoiding overlap between examples.



This generated training table is then connected to the graph representation used by RDL. 
At prediction time, the process is nearly identical, except that the label column is not computed---we only use the part of the query that defines the set of valid entities. The prediction table contains candidate rows to be predicted for the latest date in the dataset, unless a custom timestamp is provided.

Spark already comes with effective optimizations, which we generally rely on for efficient computation. We avoid triggering any Spark actions until the entire plan is generated, executing the entire plan in a single large Spark query.
Nevertheless, we find that in certain situations, Spark optimizations fall short, and a collection of high-level optimizations significantly improves the processing.

\textbf{Avoiding a timestamp and entity cross join:}
Combining a collection of timestamps with all entities is equivalent to a cross join.
In practice, using the \texttt{crossJoin} operation anywhere in the Spark plan is slow as it performs unnecessary duplication and sharding of the entity table.
To mitigate this, we exploit the fact that the number of timestamps is typically small (often fewer than $1000$).
This allows us to replace the cross join with a custom user-defined table function (UDTF) that performs the same mapping more efficiently.
This optimization is particularly effective when entity tables include validity intervals, such as user sign-up and churn dates, as invalid (entity, timestamp) pairs can be discarded directly within the UDTF (e.g., timestamps preceding a customer’s sign-up date).

\textbf{Avoiding computing labels for filtered-out entities:}
Executing the entire label computation plan in a single Spark query has a downside:
In most production queries, we are only interested in a subset of entities defined by entity filters (e.g., \texttt{\textcolor{foreach}{FOR EACH} CUSTOMER.CUSTOMER\_ID \textcolor{where}{WHERE} CUSTOMER.LOCATION\_ID='NY'}).
Ideally, all expensive operations should be restricted to this subset.
However, depending on join order, Spark may compute targets for entities that are later filtered out, or repeatedly reapply filters across the execution plan.

To reduce this overhead, we decompose the Spark plan into four stages, materializing intermediate results after each step to prevent recomputation:
\begin{itemize}
    \item \textbf{Static Entity Filters} : If a query has a static entity filter that can be applied independently of timestamps, it is applied ahead of time. E.g., if the entity definition is \\
    \texttt{\textcolor{foreach}{FOR EACH} CUSTOMERS.CUSTOMER\_ID \\ \hspace*{8pt}  \textcolor{where}{WHERE} CUSTOMERS.LOCATION\_ID = 'NY' AND \\
    \hspace*{12.2pt}\textcolor{aggr}{COUNT(}TRANSACTIONS.*\textcolor{aggr}{,-10, 0, days)} > 0},\\ 
    the static condition \texttt{CUSTOMERS.LOCATION\_ID = 'NY'} can get applied as part of this step.
    \item \textbf{Entity -- Timestamp cross-join} : All remaining entities are combined with valid timestamps in a UDTF, as introduced earlier in this section.
    \item \textbf{Temporal Filters} All filters that depend on timestamps are applied. In the example from step 1, this means applying the filter \texttt{\textcolor{aggr}{COUNT(}TRANSACTIONS.*\textcolor{aggr}{,-10, 0)} > 0}.
    \item \textbf{Target Computation} Target computation is conducted last as it is the least likely step to discard rows. Rows can still be dropped in the case of malformed data or undefined aggregations (e.g.,\ \texttt{\textcolor{aggr}{MAX}} of an empty table); however, this usually represents only a small fraction of the total rows in the final table.
\end{itemize}
\vspace{2mm} 

\xhdr{Implementation Execution Time Experiments}

\begin{table}[t]
\begin{tabular}{lllll}
Dataset & none & X-join & Filter & both \\ \hline
 rel-Amazon (small) & $2.5$ & $2.5$ & $2.5$ & $2.5$\\
 Fannie Mae (medium) &  $15$ & $12$ & $14$& $10$\\
 H\&M (large) & timeout & $115$ & timeout & $96$\\
\end{tabular}
\vspace{4mm}
\caption{
\textbf{Spark execution times}: \textnormal{Running time (in minutes) for three queries on datasets of increasing size. We compare four configurations: no optimizations, avoiding timestamp--entity cross joins (\emph{X-join}), avoiding computation for filtered-out entities (\emph{Filter}), and both optimizations combined. The results show substantial gains on medium and large datasets, with the cross-join optimization being essential to complete the large-scale experiment within the 10-hour timeout.}}
\label{table:sparktime}
\end{table}

We benchmark these optimizations on three popular relational deep learning datasets of increasing scale.
All experiments are implemented in Spark~\cite{spark} and run on the same elastic map reduce cluster.
We intentionally select queries with aggressive filters to measure the benefit of early entity pruning.
In practice, we observe that the two main contributing factors to the training table generation time are the entity table size and the number of timeframes.

The small-scale experiment runs on the rel-Amazon dataset, which contains Amazon reviews of book-related products~\cite{relbench, amazon_reviews}. The dataset contains three tables consisting of $1.8$M customers, $506k$ products, and $20M$ reviews.
The task is to predict whether the count of distinct ratings for products that have not received a rating above $3$ in the past month is non-zero:
\\ \hspace*{15pt}
\texttt{\textcolor{predict}{\textbf{PREDICT}} \textcolor{aggr}{COUNT\_DISTINCT(}\\ \hspace*{30pt}
RATINGS.RATING \textcolor{aggr}{0, 60, days} \\ \hspace*{13pt}
\textcolor{aggr}{)} \textcolor{where}{> 0}\\ \hspace*{13pt}
\textcolor{foreach}{\textbf{FOR EACH}} PRODUCTS.PRODUCT\_ID\\ \hspace*{13pt}
\textcolor{where}{\textbf{WHERE}} \textcolor{aggr}{MAX(}RATINGS.RATING\textcolor{aggr}{, -30, 0, days)} \textcolor{where}{< 3}
} \\
The training table spans $88$ time frames; however, only $1573$ entities meet the entity filter.

Our medium-scale experiment runs on the Fannie Mae credit risk dataset\footnote{Available at \url{https://capitalmarkets.fanniemae.com/credit-risk-transfer/single-family-credit-risk-transfer/fannie-mae-single-family-loan-performance-data}}.
The medium-scale experiment uses the Fannie Mae credit risk dataset, containing $15$M loans and $600$M payments. The task predicts the total sum of credit payments over the next $60$ days for loans with a single borrower:
\\ \hspace*{15pt}
\texttt{\textcolor{predict}{\textbf{PREDICT}} \textcolor{aggr}{SUM(} \\ \hspace*{30pt}
PAYMENT\_HISTORY.MONTHLY\_PAYMENT \textcolor{aggr}{0, 60, days}\\ \hspace*{13pt}
\textcolor{aggr}{)}\\ \hspace*{13pt}
\textcolor{foreach}{\textbf{FOR EACH}} DIM\_LOANS.LOAN\_IDENTIFIER\\ \hspace*{13pt}
\textcolor{where}{\textbf{WHERE}} DIM\_LOANS.NUMBER\_OF\_BORROWERS \textcolor{where}{<= 1}
} \\
The query produces $25$ timeframes of data, with $4$M entities meeting the filter condition.

Finally, we run a large-scale experiment on a synthetically-upscaled variant of the H\&M e-commerce dataset~\cite{relbench}. 
The dataset usually contains three tables that describe $1.5$M customers, $31$M transactions, and $105$k articles, but we scale it to $275$M customers and $6.3$B transactions by duplicating the entries of the customer and transaction tables. The task predicts the count of transactions in the next $7$ days for all customers over the age of $30$:
\\ \hspace*{15pt}
\texttt{\textcolor{predict}{\textbf{PREDICT}} \textcolor{aggr}{COUNT(}TRANSACTIONS.* \textcolor{aggr}{0, 7, days}\textcolor{aggr}{)}\\ \hspace*{13pt}
\textcolor{foreach}{\textbf{FOR EACH}} CUSTOMERS.CUSTOMER\_ID\\ \hspace*{13pt}
\textcolor{where}{\textbf{WHERE}} CUSTOMERS.AGE \textcolor{where}{> 30}
} \\
The query produces $70$ timeframes of data with $143$M entities that meet the filter conditions.

Table~\ref{table:sparktime} summarizes the execution times and highlights the necessity of optimization at scale: the unoptimized configuration fails to complete within the 10-hour limit on the largest dataset, whereas optimized variants succeed. As expected, performance gains are negligible for small datasets.

The benefits of these optimizations are twofold.
First, they achieve a speedup of over 6 times or more on queries that would otherwise require extensive resources, depending on the filters and scale.
Second, they allow the user to reliably speed up the label generation by adding additional filters, e.g.\ by using only a subset of entities.

\begin{figure}[t]
    \centering
    \includegraphics[width=1\linewidth]{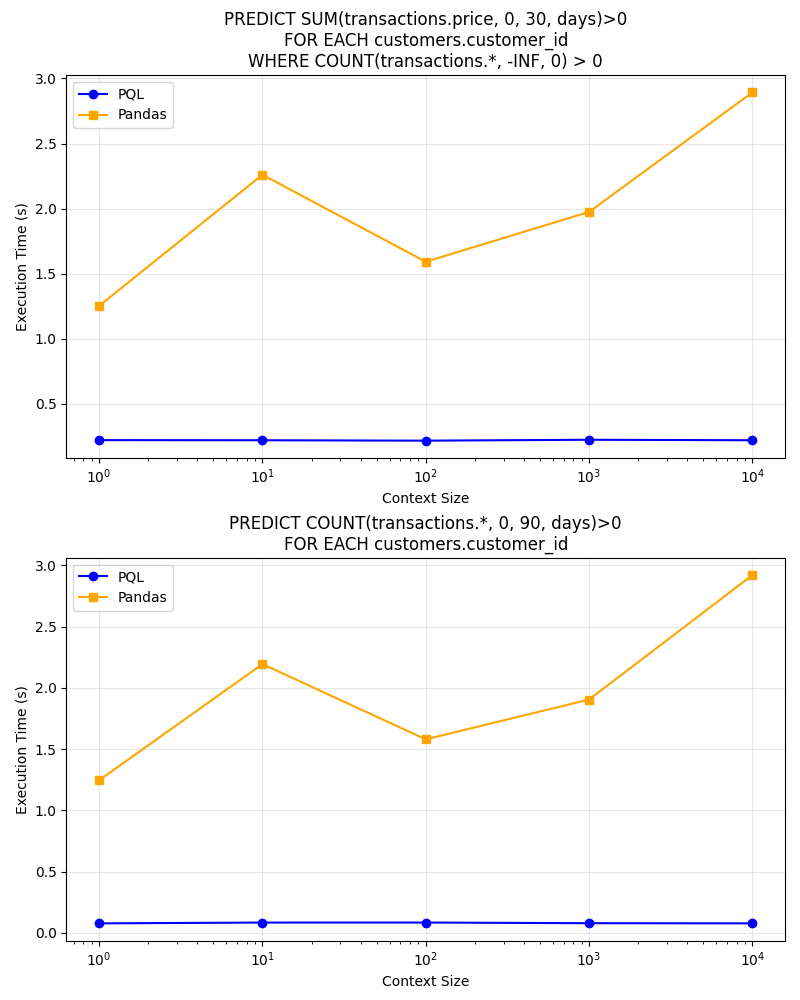}
    \caption{\textbf{PQL for RFM execution time.} \textnormal{Comparison of PQL execution time for different context sizes, compared to a naive Pandas implementation. Experiments were run on the H\&M dataset~\cite{relbench}. All running times were averaged over 10 random seeds and do not involve data reading. PQL implementation is significantly faster and more consistent, due to its ability to efficiently access the relevant subset of the database. With an efficient mechanism for data access, the size of the context does not impact the run time.}}
    \label{fig:RFM_execution_time}
\end{figure}

\subsection{Usecase: Relational Foundation Model}
The implementation of PQL for Relational Foundation Models differs substantially from its batch processing counterpart.
The Relational Foundation Model~\cite{kumorfm} is a large pre-trained model for relational deep learning that can generate predictions in seconds.
The speed-up comes from the fact that this model does not perform any traditional parameter-tuning and instead extrapolates directly from context examples provided as part of the input.

The typical context window of a Relational Foundation Model contains around $10,000$ examples, and the model is typically used interactively.
Unlike in the RDL setting, we do not try to generate data from all possible combinations of users and timestamps.
Instead, the context window contains a collection of the most relevant entities for the most recent timestamps.
The key to achieving low-latency label computation is an efficient sampling implementation that, given the list of entity IDs, collects exactly the required rows from the otherwise large database.

To achieve such efficient sampling, we take advantage of database pre-processing, performed as part of data pre-processing for the Relational Foundation Model.
During this step, the database is transformed into a heterogeneous graph: rows become nodes, and primary--foreign key relationships become edges.
The PQL sampler operates on this precomputed graph.

ecause all aggregations and filters follow connections defined in the schema, the rows required to compute a query form a connected subgraph of the heterogeneous graph.
This subgraph can be efficiently collected with a custom Python module with a C++ backend based on the GraphSAGE algorithm~\cite{graphsage,fey2025pyg20scalablelearning}. 
Compared to a standard GraphSAGE sampler, the PQL sampler differs in two key ways.
First, the sampler does not sample all node and edge types, but only the types used in the label computation.
Second, unlike the typical GraphSAGE sampler, which samples a fixed number of neighbors at each step, the PQL sampler samples all neighbors within a specified time range.

Once the relevant rows of the database have been extracted, the execution of the logical plan from Section~\ref{section:implementation} is executed in Pandas~\cite{reback2020pandas}.
In practice, the total amount of data at this stage typically does not exceed a few megabytes, requiring no sophisticated plan optimizations for efficient execution.
In practical experiments, we find that label computation takes a fraction of a second---latency that is acceptable for an interactive user application but not achieved with a naive implementation.

\xhdr{Implementation Execution Time Experiments}
To assess the efficiency of the implementation, we compare the Relational Foundation Model implementation of PQL to a naive Pandas implementation.
We benchmark the two on the H\&M e-commerce dataset~\cite{relbench} which contains 2 years of clothes purchases at a major retailer with $1.3M$ customers, $105k$ articles, and $31M$ transactions.
The implementations are compared on two queries; predicting the count of transactions in the next month and predicting the spend of all customers with at least one historical transaction.
Unlike in the Relational Deep Learning implementation, we find that the number of timeframes and the size of the entity table do not significantly affect the execution time, mostly due to the small number of generated context examples.
Instead, the main bottleneck is caused by the number of operations required to execute the query.
To this end, we choose queries with a larger aggregation window.

The execution times are plotted in Figure~\ref{fig:RFM_execution_time}, demonstrating up to a 40-fold faster execution of our custom PQL implementation.
The faster runtime stems from the ability to efficiently access the relevant rows of the database, avoiding the need for time-consuming joins.
Using a custom PQL engine for the relational foundation model is thus essential for achieving sub 1 second latency, which is required for an interactive session.

\section{Discussion and Conclusion}
This paper introduces Predictive Query Language, a novel language for predictive modeling on relational databases.
The syntax supports versatile uses across a collection of real-world predictive models and can be implemented both at-scale and in a low-latency regime.
At the time of writing, it is underpinning predictions within multiple popular applications, including social networks, food delivery services, and financial institutions.
Future expected extensions of PQL will aim to increase the expressivity of the language through a collection of features.
While we find that the existing syntax covers the majority of practical use cases, we aim to reduce the amount of pre-processing required for the more involved queries. The limitations of this language can be split into two groups: the fundamental limitations arising from the assumptions and the missing functionality that we aim to add in future iterations of the language.

Through practical use of PQL on multiple real-world databases we found that some of them do not follow best practices of database construction on which the assumptions in Section~\ref{sec:assumptions} were built.
For example, if records are inserted without the appropriate timestamp annotation or are modified retroactively without an appropriate timestamped record, the procedure of generating training data from the past records is bound to produce inaccurate labels.
In majority of such cases, we found that additional manual pre-processing and data cleaning can make PQL and RDL applicable.
Unfortunately, such pre-processing does not only present additional labor, but also a model accuracy reduction as otherwise valuable data needs to be avoided to avoid temporal information leakage.

Even though the existing PQL syntax covers most of the problems we have encountered in practical applications, we find that additional syntax features could further improve its usability:
\begin{itemize}
    \item \textbf{Explicit Join Support}: Explicit syntax for data joins would allow queries where label is computed from data that does not sit in a table directly linked to the entity table. Moreover, it provides more control in cases where two tables are connected with more than one foreign key and either can be used for aggregations.
    \item \textbf{Operations between Columns and Aggregations:} In the described syntax, only comparisons between a column/aggregation and a constant is permitted. Addressing this limitation is planned for future iterations of PQL.
    \item \textbf{Non-Atomic Primary Key Support}: In the current language design, each prediction corresponds to exactly one entity, represented as a primary key. Certain usecases do not fit within this assumption, e.g.\ when a prediction is made for a pair or a set of entities. These require additional pre-processing where said pairs or sets are introduced as an auxiliary table. This effort could be avoided through an enhanced syntax in the \textcolor{predict}{\texttt{PREDICT}} clause.
\end{itemize}

The ongoing deployments and active users of PQL demonstrate that, combined with Relational Deep Learning and Relational Foundation Models, it unlocks rapid and high-quality predictive model development, paving the way to faster development and novel use cases.

\xhdr{Acknowledgments}
We would like to thank Kumo.ai team for their invaluable support. Special thanks go to Min Shen, Alex Porter, Dong Wang, and Subramanya Dulloor for support on technical challenges, Federico L{\'o}pez and Valter Hudovernik for reviewing the manuscript, and Myunghwan Kim, Zack Drach, Tin-Yun Ho, and Vanja Josifovski for their helpful suggestions in language design.

\bibliographystyle{ACM-Reference-Format}
\bibliography{sample}


\end{document}